\documentstyle[eqsecnum,multicol,psfig,aps,prl,array]{revtex}
\newcommand{\bleq}{\ifpreprintsty
		   \else
		   \end{multicols}\widetext \vspace*{-3.5ex}{\tiny
		   
		\noindent\begin{tabular}[t]{c|}
		   \parbox{0.493\hsize}{~} \\ \hline \end{tabular}}
				      \fi}
\newcommand{\eleq}{\ifpreprintsty
		   \else
		   {\tiny\hspace*{\fill}\begin{tabular}[t]{|c}\hline
		    \parbox{0.49\hsize}{~} \\
		    \end{tabular}}\vspace*{-2.5ex}\begin{multicols}{2}
		    \narrowtext
		    \fi}
\newcommand{\bcols}{\ifpreprintsty\else\begin{multicols}{2} 
	\narrowtext\fi}
\newcommand{\ecols}{\ifpreprintsty\else\end{multicols}\fi}

\draft
\widetext
\begin{document}
\title{Nucleation model for the description of glass formation} 
\author{Matthieu Micoulaut}
\address{Laboratoire GCR-CNRS URA 769\\
Universit\'e Pierre et Marie Curie, Tour 22, Boite 142\\
4, Place Jussieu, 75252 Paris Cedex 05, France\\}

\date{\today}
\maketitle
\begin{abstract}
\par
We present in this letter a model of glass formation using energy barriers
and a crystal nucleation process. We then analyze the corresponding dynamical
equation in the vicinity of the stationary solutions. The occurence of a pure
amorphous solution (i.e. glass) is due to the contribution of a ratio $\Lambda$ depending
on the cooling rate and the crystal nucleation frequency. We also construct
time-temperature transformation (T-T-T) curves in order to relate the model
with the kinetic treatment of glass formation.
\par
{Pacs:} 61.20N-81.20P
\end{abstract}
\bcols 
The formation of glasses requires cooling to a sufficiently low temperature,
below the glass transition temperature, without the occurence of detectable 
crystallization.
\par
One of the most comprehensive theory of dynamical processes in glasses is
the so-called mode-coupling theory (MCT) , developed by Goetze and others
\cite{r1}. The
MCT is based on a set of non-linear damped second-order differential 
equations, which couple the density correlation functions of a supercooled 
liquid in a retarded way. These equations have several interesting 
properties, such as the possibility of a stretched-exponential time decay
for the correlation function for particular values of the strength of 
the non-linear coupling. They exhibiti, also, a kind of "{\em ideal glass}"
transition, at an ideal glass temperature $T_C$, corresponding to a 
singularity of the equations. 
Nevertheless, the major limitation is that, so far, only spherically symmetric 
interatomic interactions have been considered, as for instance the computer
calculation by Bengtzelius on a Lennard-Jones system \cite{r2}. Thus, 
the complexities in real glass-forming systems arising from covalent interactions (e.g.
silica) are neglected, so that the theory seems only well-adapted for 
fragile glass-forming liquids (e.g. organic or molecular glasses). 
There have been recently also theoretical developments of mean-field models in order to
understand the role of different factors in the glass transition, such as
configurational entropy \cite{r3}, energy barriers \cite{r4} or free energy
\cite{r5}, some of them being inspired from spin glass models. However, the
critical question in discussing glass formation is not whether the solid
will be obtained from a melt quench, but how fast a given liquid must be
cooled in order to prevent the kinetic processes involved in crystallization. 
Thus, the cooling rate and the crystal nucleation rate should play the major 
role. 
\par
The work presented here consists in the construction of an exactly solvable 
mean-field model of crystalline nucleation, based on a probabilistic 
description, which can be described by a dynamical equation. 
Although this model is rather abstract and has no thermodynamic transition, 
we believe that it captures at least
parts of the physics involved in the glass transition and explains it in a
simple fashion. The model attempts also to relate factors that are 
widely viewed as decisive in the formation and preparation of glasses, namely 
structural, kinetic and thermodynamic factors \cite{r6}.
We shall see that the probability of occurence of an 
amorphous phase (identified with a supercooled liquid below the melting
temperature and then with a glass when temperature has decreased 
enough) depends strongly on the cooling rate.
Let us try to describe what happens on a microscopic scale
when a liquid is cooled and translate this in terms of
thermodynamics. 
\par
A liquid at high temperature (with T higher than the melting 
temperature) should contain a lot of aggregates of different sizes 
and shapes. The atoms which compose some of the aggregates may be already 
organized with periodic character (identified with a crystalline nucleus), 
whereas others do not exhibit this feature and may display a disordered structural 
organization (amorphous structure).\par 
{\em Construction of the model.} Let $p_c=p({\cal C}_0)$ be the probability 
of finding an atom, in a crystalline aggregate, being in the initial liquid 
configuration ${\cal C}_0$. 
The structure associated with this aggregate is size-independent and there
is no reference to a spatial structure, therefore the model is of mean-field
type. In this description, the probability $p_c$ corresponds to the 
number of atoms which are already trapped inside a crystalline unit cell or 
a set of such cells, over the total number of atoms. Similarly, we define 
the probability of finding an atom inside an amorphous aggregate by $1-p_c$. 
The glassy state will thus correspond to $p_c=0$ at low temperature and the
crystal to $p_c=1$.
The probability $p_c$ depends at least on time, temperature and the cooling rate.
When the temperature is decreasing, the aggregates start to grow 
by nucleation. The energy barriers related to such agglomeration processes
depend, of course, on the nature of the clusters which are involved. 
Therefore, we define for a crystal-crystal nucleation an energy barrier of $2E_c$, for
a crystal-amorphous structure agglomeration an energy barrier of $E_c+E_a$
and for an amorphous-amorphous structure agglomeration $2E_a$. 
$E_c$ is the mean energy stored in a crystalline cluster and $E_a$ the one 
stored in an amorphous structure. 
This means that the initial configuration ${\cal C}_0$ can be considered
as a system with two energy states $E_c$ and $E_a$ and that the atoms 
can be treated as particles able to occupy one of these states.
After agglomeration or nucleation, the system is in a final configuration 
${\cal C}$.
It is composed of three energy states ($2E_c$, $E_a+E_c$ and $2E_a$), and the 
repartition of the particles can be computed. 
The probability of finding an atom trapped in such a new state may be 
proportional to the product of the initial probabilities (i.e. to the number
of related particles) and a Boltzmann weight, 
inspired from the Thouless-Anderson-Palmer (TAP) description 
\cite{r7} of spin glass theory (e.g. the
pure crystal state with probability $p_{cc}$ is proportional to $p_c^2$). Such a
description has been also used in a growth model of polygons \cite{r8} and in
structural glass models \cite{r8b}.
\par
\begin{eqnarray}
\label{1}
p_{ij}\ =\ {\frac {(2-\delta_{ij})}{{\cal Z}}}(1-p_c)^{(\delta_{ia}+\delta_{ja})}
p_c^{(\delta_{jc}+\delta_{ic})}e^{-\alpha(\delta_{ic}+\delta_{jc})} \nonumber
\end{eqnarray}
where $\alpha =\beta(E_c-E_a)$ with $\beta$ the reverse temperature. Since 
the probabilities are all normalized by the factor ${\cal Z}$, only the 
difference
$\alpha$ is relevant. The normalizing factor ${\cal Z}$, which can be  
regarded as a TAP-like partition sum ensures that $\sum_{i,j}p_{ij}=1$, as it 
should be. 
During nucleation, an atom trapped inside a {\em crystalline} state, can be
found in the nucleated configuration $\cal C$ (energy states $2E_c$ or $E_c+
E_a$) or still in the initial configuration ${\cal C}_0$ (energy state $E_c$).
Its probability is expressed by:
\begin{eqnarray}
\label{5}
p_c(t)\ =\ \biggl[1\ -\ s(t)\biggr]\ p({\cal C}_0)\ +\ s(t)\ p({\cal C}) \nonumber
\end{eqnarray}
where:
\begin{eqnarray}
\label{6}
p({\cal C})\ =\ {\frac {1}{2}}\biggl[2\ p_{cc}\ +\ p_{ac}\biggr] \nonumber
\end{eqnarray}
The function $s(t)$ with $0\leq s(t)\leq 1$, introduced above depends on time
and represents the growth rate (i.e.
a measure of the number of nucleated crystalline aggregates which are already 
formed). Its derivative ${\dot s}(t)=ds/dt$ is then the nucleation frequency 
or growth
speed. When $s(t)=1$,
the nucleation process is finished. 
\par
The derivative of the probability $p_c(t)$ obeys the following dynamical 
equation, which is constructed from the equations above and which takes into 
account the effect of the cooling rate $Q=-dT/dt$.
\begin{eqnarray}
\label{7}
{\frac {dp_c(t)}{dt}}\ =\ {\dot s}(t)\biggl[p({\cal C})\ -\ p({\cal C}_0)
\biggr]\ -\
Q\ {\frac {\partial p({\cal C})}{\partial T}} \nonumber
\end{eqnarray}
At finite temperature and for long enough times, the model reaches a stationary
state, corresponding to thermal equilibrium, thus we 
shall be interested in the stationary solutions of the dynamical equation.
We can solve this equation in terms of the probability $p({\cal C}_0)$, 
since the
probability $p({\cal C})$ of the final state is also constructed with them. 
This should happen in the liquid and supercooled state for very slow
cooling rates and in the final solid for $T\simeq 0$.
Obvious solutions of the equation are $p({\cal C}_0)=0$ (pure amorphous state) 
and $p({\cal C}_0)=1$ (pure crystal state). Besides solutions without physical 
meaning ($p({\cal C}_0)>1)$, an intermediate solution exists which shifts with the
cooling rate $Q$, the crystal nucleation frequency ${\dot s}(t)$ and the 
temperature:
\begin{eqnarray}
\label{9}
p_c^{int}\ =\ {\frac {e^u\biggl[2-{\frac {\Lambda}{T}}u+{\frac 
{\Lambda}{T}}u\coth u\biggr]}{4 \sinh u}}  \nonumber
\end{eqnarray}
where $\Lambda =Q/{\dot s}(t)$ and $u=\alpha /2$. The existence of this intermediate solution
depends on $\Lambda$, which must be inside region II of fig. 1.\par 
{\em Dynamics of crystal and glass formation}. 
In order to see in which direction and under what condition the system 
can evolve, we have performed the linearization of the dynamical equation in 
the vicinity of the stationary solutions and computed the corresponding
relaxation times $\tau_0 (T)$ and $\tau _1 (T)$, which characterize the
convergence of the nucleation process towards equilibrium. The behavior of 
the liquid is mainly driven by the factor $\Lambda$ and typical 
situations occur, according the sign of $\alpha$.\newline
\begin{figure}
\begin{center}
\psfig{figure=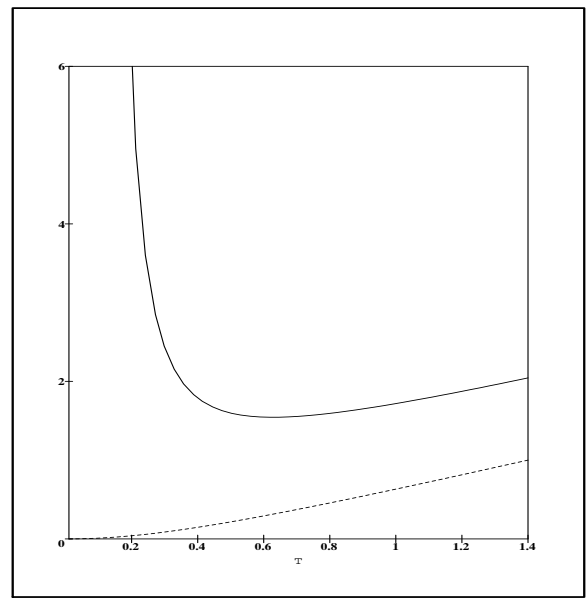,width=\linewidth,height=\linewidth}
\caption{A plot of the critical factors $\vert \Lambda_1\vert $ (solid line),
which yields $p_c^{int}=1$,   
and $\vert \Lambda_2\vert$ (dashed line), which yields $p_c^{int}=0$, as a 
function of temperature $T$
with $E_c-E_a=-1$. The intermediate solution $p_c^{int}$ exists only if 
$\vert \Lambda_1\vert
<\vert \Lambda\vert <\vert \Lambda 2\vert$.}
\end{center}
\end{figure}
When $\alpha <0$ (the crystalline energy barrier being lower than the amorphous
one), there are only two solutions
and the $p({\cal C}_0)=1$ one is an attractor. This means that every 
fluctuation $\xi$ in the probability which could occur during the nucleation 
process, tends to vanish with increasing time. The dynamical equation becomes: 
$\tau_1 (T)d\xi=-\xi dt$ with relaxation time:
$$\tau_1(T)\ =\ {\frac {\tanh {\frac {\alpha}{2}}-1}{({\frac {\alpha \Lambda}
{T}}+4)\tanh {\frac {\alpha}{2}}+{\frac {\alpha \Lambda}{T}}}}$$
Growth is then made only out of atoms being part of crystalline aggregates.
On the contrary, it is easy to verify that a fluctuation $\eta$ in the 
vicinity of the amorphous solution $p({\cal C}_0)=0$ will grow exponentially 
with increasing time and the relaxation time $\tau_0 (T)$ [$\tau_0 
(T)d\eta=\eta dt$]:
$$\tau_0(T)\ =\ {\frac {1+\tanh {\frac {\alpha}{2}}}{({\frac {\alpha \Lambda}
{T}}-4)\tanh {\frac {\alpha}{2}}-{\frac {\alpha \Lambda}{T}}}}$$
So, there is no dissipation of an amorphous fluctuation $\eta$.
We believe that this situation is typical of systems for which there is no
possibility in glass formation (e.g. all the elements except selenium, sulfur
and tellurium).
\begin{figure}
\begin{center}
\psfig{figure=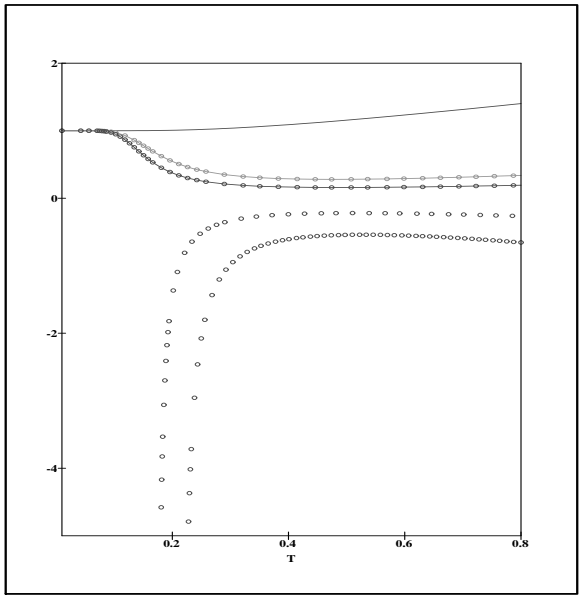,width=\linewidth,height=\linewidth}
\caption{The relaxation time $\tau_0(T)$ as a function of 
temperature for two different processes (cooling [circles] and 
heating [circles and solid line]) and two different values of 
$\vert \Lambda\vert$ ($5,10$). The solid line corresponds to 
$\tau_{equ}$.}
\end{center}
\end{figure}
An interesting feature of this dynamical analysis is given by the relaxation
times $\tau_0 (T)$ and $\tau_1(T)$ which exhibit a very different behavior 
following the process (cooling or heating). 
We define by $\tau_{equ}$ the relaxation time of a process at thermal equilibrium
(i.e. when $\Lambda$ is negligible, so $1/\tau_{equ}=1-e^{\alpha}$. For
low temperatures ($T\rightarrow 0$), the dynamics of the process is close to
equilibrium since $\tau_1(T)\simeq \tau_{equ}$. Starting from a high T liquid,
one observes that $\tau_1$ decreases when temperature goes down. This means
that it becomes more and more easier for the system to fall on the crystalline
attractor. $\tau_1$ exhibits a minimum at a certain value $T^*=-\Lambda+
\sqrt{\Lambda^2-\Lambda (E_c-E_a)}$, which we identify with the melting
temperature. It is the temperature at which the driving force towards the
crystalline aggregation ($p({\cal C}_0)=1$) is at its maximum. In the heating
process ($\Lambda <0$), the $p({\cal C}_0)=0$ and $p({\cal C}_0)=1$ solutions
are both attractive, the latter with a relaxation time $\tau_1\simeq \tau_{equ}$
for low T, but this solution diverges rapidly in the vicinity of a temperature
$T_1$ close to $T^*$. For $T>T_1$, the crystal solution is unstable and the
amorphous (liquid) solution only is attrative. In the 
range $0<T<T_1$, the intermediate solution exists but it is unstable. The 
described dynamical behavior shows that a crystalline solid which is
heated from a low temperature, remains the preferential structure up to a
temperature $T_1$ close to the temperature $T^*$. There, the 
crystal relaxation time $\tau_1$ diverges, which allows the possibility for the 
system to investigate other possible structural pathways, such as the 
amorphous one. 
\par
When $\alpha>0$, there is a possibility of formation of an amorphous
solid, because the $p({\cal C}_0)=0$ solution is attractive, whereas, the
crystalline state is repulsive. In this situation, $\tau_0(T)$ displays the
following behavior: starting from a high $T$ liquid phase, one can easily 
check that the fluctuations in the vicinity of the crystalline solution grow
exponentially with a relaxation time $\tau_1(T)$. The $\tau_0(T)$ relaxation
time is negative and close to 0. It starts to decrease when the temperature 
decreases (fig. 2, circles), i.e. the dynamics of the attractor becomes
slower. the liquid state is still the preferential one, but the time needed
for a created fluctuation to vanish, increases dramatically. At some
temperature $T(\Lambda )$, related to the factor $\Lambda$, we have 
$\tau_0(T)\rightarrow -\infty$. We identify
this temperature with the glass transition temperature $T_g$, which depends
on the cooling rate, as it should do. The configurational change $\eta$
which may cause the relaxation for the low-temperature supercooled liquid
has become infinitely slow, thus the liquid behaves as a solid, and $T_g$
satisfies:
$$\tanh \biggl[{\frac {E_c-E_a}{2T_g}}\biggr]\ =\ {\frac {\Lambda (E_c-E_a)}{4T_g^2-\Lambda (E_c-E_a)}}$$
On the contrary, we can remark that starting from $T=0$ (at equilibrium)
$\tau_0(T)$ is close to $\tau_{equ}$ and the $p({\cal C}_0)=0$ solution is
now repulsive (fig. 2, solid line with circles). $\tau_0(T)$ rapidly falls 
to 0, so the configurational change is possible and crystalline fluctuations 
can grow 
easily, in agreement with current observation stating that recrystallisation
occurs when a glass is heated up.\par
{\em Construction of the time-temperature-transformation (T-T-T) curves.} As
we have constructed this model by using a simple crystallization process, it 
is interesting to relate it to results concerning the kinetic treatment of 
glass formation, which describe the crystallization process by considering both
nucleation and crystal growth and estimate the cooling rates required to
form glasses \cite{r6}. Thus, kinetic treatments of glass formation are based on 
identfying a certain value of the volume fraction crystallized $v_c/v$, as
borderline between an amorphous and a crystalline solid \cite{r9} ($v_c/v$ is 
generally of the order of the experimentally just detectable degree of 
crystallinity, i.e. around $10^{-6}$). The results of such investigations are 
plotted on a $t-T$ plane for different values of $v_c/v$ and represent the
time-temperature-transformation (T-T-T) curves \cite{r6}.
\begin{figure}
\begin{center}
\psfig{figure=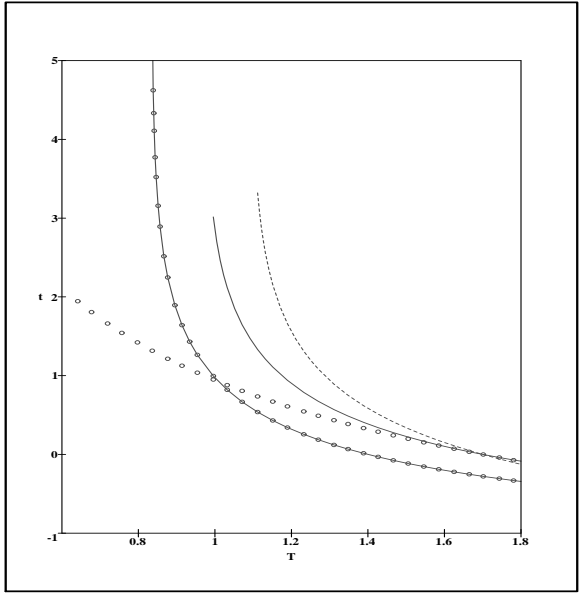,width=\linewidth,height=\linewidth}
\caption{The T-T-T curves for different s(t) behavior a) $s(t)=1-e^{-t}$.
Solid line (with points) $T_0=1.4$, solid line $T_0=1.7$, both with a fraction 
of $x_c=10^{-6}$. Dashed line: $T_0=1.7$ and $x_c=0.2$. b) $s(t)=t$. Dots: 
system with $T_0=1.7$, $x_c=10^{-6}$. As before, $E_c-E_a=-1$ for all 
situations.}
\end{center}
\end{figure}
In our mean-field description, the extensive variables $p_c$ and $v_c$
are related by: $v_c=p_c\ v$, where $v$ is the total volume of the system.
Therefore, if $v_c=p_c v$ is maintained constant, the right hand side of the
dynamical equation will be 
equal to 0 for any temperature. We can then compute the T-dependence of the
factor $\Lambda$ and obtain after integration a $T-t$ relationship.
If $x_c$ is the fraction of crystalline atoms (with probability $p_c^{int}$), 
then $\Lambda$ must satisfy:
$$x_c(e^{\alpha}-1)^2-e^{2\alpha}+e^{\alpha}=-{\frac {\Lambda}{T}}\alpha e^{\alpha}=
-{\frac {\alpha e^{\alpha}}{T}}\ {\frac {dT}{{\dot s}(t)dt}}$$
and integration yields:
$$\alpha\biggl({\frac {T}{T_0}}-1\biggr)+ln\ \biggl[{\frac {\alpha(x_c-1)-x_c}
{\alpha_0(x_c-1)-x_c}}\biggr]=s(t)$$
where $T_0$ is the initial temperature of the melt and $\alpha_0$ the corresponding 
factor $\alpha$. We have plotted different situations in fig. 3 for different
t-dependences of the function s(t). One natural (and the simplest) dependence 
for $s(t)$ is the exponential one with $s(0)=0$ and $s(\infty)=1$, in order
to agree with the construction of $p_c(t)$. Figure 3 shows the dependence of
the temperature with time if one keeps $x_c$ constant during nucleation.
The curves displayed show the same behavior as the constructed T-T-T
curves by Uhlmann \cite{r6,r9}. i) The T-t curve with a lower degree of 
crystallinity ($x_c=10^{-6}$) envelopes the curve with a higher degree 
($x_c=0.2$), i.e. the temperature must decrease more rapidly, as seen on 
the figure. ii) The effect of the initial temperature 
$T_0$ can also be observed. 
\par
In conclusion, we should stress that the dynamical analysis presented here 
puts forward the general accepted picture that a glass is obtained by cooling 
a melt enough in order to avoid nucleation, although this picture displays no
transition in the thermodynamical sense. However, different physical
parameters are involved in this very complex transition such as 
free energy, cooling rate, nucleation frequency, initial temperature or 
structure. The work presented here was an attempt for the description
of the glass formation using a nucleation process. Forthcoming work will 
include structural factors \cite{r10}.

\ecols
\end{document}